\documentclass{sf2a-conf}
\usepackage{graphicx}

\begin{document}
\TitreGlobal{SF2A 2007}
\title{Visible and invisible molecular gas in collisional debris of galaxies}


\author{P.-A.  Duc} \address{AIM, DSM/CEA -- CNRS -- Universit\'e Paris Diderot, DAPNIA/Service d'astrophysique, CEA-Saclay, 91191 Gif sur Yvette cedex, France}

\runningtitle{Molecular gas in collisional debris}

\setcounter{page}{1}

\index{Duc P.-A.}

\maketitle

\begin{abstract}
Molecular gas has been searched for and found in unexpectedly large quantities in some collisional debris of interacting galaxies: HI-rich tidal tails, bridges and collisional rings. It was so far observed through  millimeter  observations of the CO line  and detected towards or near regions of star-formation associated to dense condensations of the atomic hydrogen. The discovery of cool H$_{2}$ at distances greater than 50 kpc from the parent (colliding) galaxies, whereas the external disk of spirals is generally considered to be CO--poor,  raised question on its origin and favored the hypothesis of a  local production out of collapsed HI clouds. However recent  observations of a diffuse CO component along tidal debris have challenged this idea. 
Another recent puzzle is the measurement in the collisional debris of  two interacting systems and four recycled objects of a missing mass, whereas no dark matter is expected there.  One debated interpretation is that this unseen component is cold, ``invisible"  molecular gas initially present in the disk of spirals. 
\end{abstract}

%

\section{Introduction: the many components of collisional debris}

The collisional debris addressed here  refer to all the material that is expelled into the intergalactic medium during galaxy--galaxy interactions. This is the result of either the tidal forces that shape bridges and tails, of direct high-speed impacts at the origin of rings or head-on collisions forming  systems similar to the so--called ``Taffy" galaxies (see Struck 1999 for a review). Collisional debris may consist of old and young stars,  gas clouds and dust with a relative proportion that depends very much on the type of interactions and properties of the parent galaxies. Tidal tails with a prominent old stellar population will be common in slow encounters between late-type galaxies -- see the famous Antennae galaxies. Purely stellar tails may exist, when the gaseous counterpart  has for some reason been displaced and is now offset (Mihos 2001). More commonly, the gaseous tail extends further than the stellar one, because it was originally more extended in the disk of its parent galaxies. Collisional rings are particularly gas-rich, being embedded in a faint stellar halo. Finally, long, purely gaseous, tidal tails may form during high-speed collisions, as recently shown by Duc \& Bournaud (2007). Finally, if the gas in the collisional debris is dense enough, it collapses and locally  form a new generation of stars. 

Whereas stellar debris have been known for a long time -- they are the peculiarities  in the optical catalog of Peculiar Galaxies by Arp (1966) --,  their gaseous, HI, counterparts have been systematically studied from less than twenty years (e.g., Hibbard \& van Gorkom  1996). This was possible thanks to interferometers like the Very Large Array which offer a very large field of view. In situ star formation in collisional debris has been discovered through several ways, including broad-band optical (e.g. Schombert et al. 1990) and UV (e.g., Neff et al. 2005; Boquien et al., 2007) imaging, narrow-band H$\alpha$ images (e.g. Iglesias-P{\'a}ramo  \& V{\'{\i}}lchez  2001) as well as mid-infrared photometry.  If ongoing star-formation is present in collisional debris, molecular gas should be present as well. This paper reviews the various attempts that have been done so far to detect it.
 
\section{Detecting conventional molecular gas}

Cool H$_{2}$ gas is usually detected through the emission of the CO millimeter line. The abundance of carbon monoxide depends, among other factors, on the metallicity of the medium. Since tidal debris are composed of pre-enriched material, they have an oxygen abundance of about one third to half solar which  makes the detectability of the CO line relatively easy. First attempts to find it in tidal tails with the NRAO 12m antenna had nevertheless failed (Smith \& Higdon 994), with the exception of  NGC~2782 where Smith et al. (1999) detected CO at the base of one of its tidal tails. Besides,  detections of CO were reported in a few, somehow isolated, clouds, which are likely tidal debris (Combes et al. 1988; Brouillet et al. 1992; Walter \& Heithausen 1999). The first observations of molecular gas far away from the parent galaxies were made by Braine et al. (2000, 2001) who used the IRAM 30m antenna. They obtained a CO signal towards Tidal Dwarf Galaxy candidates located near the end of long, HI-rich, tidal tails.  Unexpectedly  large quantities of molecular gas  -- the equivalent of the whole Milky-Way content -- were mapped by Lisenfeld et al. (2002) on one of the tidal tails of the Stephan's Quintet compact group of galaxies.  Similarly large quantities of H$_{2}$ were also detected in Taffy-like systems in the bridge between the two colliding galaxies (Braine et al. 2003, 2004). Whenever a tidal interaction is involved, CO is detected within HI structures at locations  close to the local peak of the column density, where ongoing star formation is also observed. Besides the HI and CO lines seem to share the same kinematics. This lead Braine et al. (2000) to claim that the molecular clouds were formed locally out of collapsed dense HI clouds, instead of having been directly expelled from  the parent galaxies together with the atomic hydrogen. However recent observations make this interpretation less secure. First of all, CO has been found in gaseous debris located outside the regions of current star formation (Lisenfeld et al. 2007), in particular at location where the HI seems not to be gravitationally bound (Duc et al. 2007). Besides, CO has also been found in the outermost regions of spirals (Braine \& Herpin, 2004) whereas it was initially thought to be more concentrated towards the central regions, and thus not end up in tidal debris. Note however that wherever SF takes place, the CO lines are narrower. Thus whereas a local production of  H$_{2}$ is not excluded, especially since the time-scale for the transformation of HI into H$_{2}$ is rapid, tidal debris may also contain a more diffuse component coming from the parent disk. 
 
 Having detailed maps of the molecular gas should help to disentangle these various components. Only a few of them are yet available (Walter et al. 2006, Lisenfeld et al. 2004). As shown in Figure~1, the core of the CO emission in the TDG NGC 2992N seems slightly displaced by 0.5--1 kpc from the HI peak and SF center. In the TDG SQ-B, the CO emission follows the H$\alpha$ distribution, dust lane and mid--infrared emission, but is offset by 2-4 kpc from the HI peak. In all the systems observed, the interferometric observations only recover a tiny fraction of the CO emission detected by a single dish, indicating that  a diffuse component has been missed.

\begin{figure}[h]  
\centering 
\includegraphics[height=11cm]{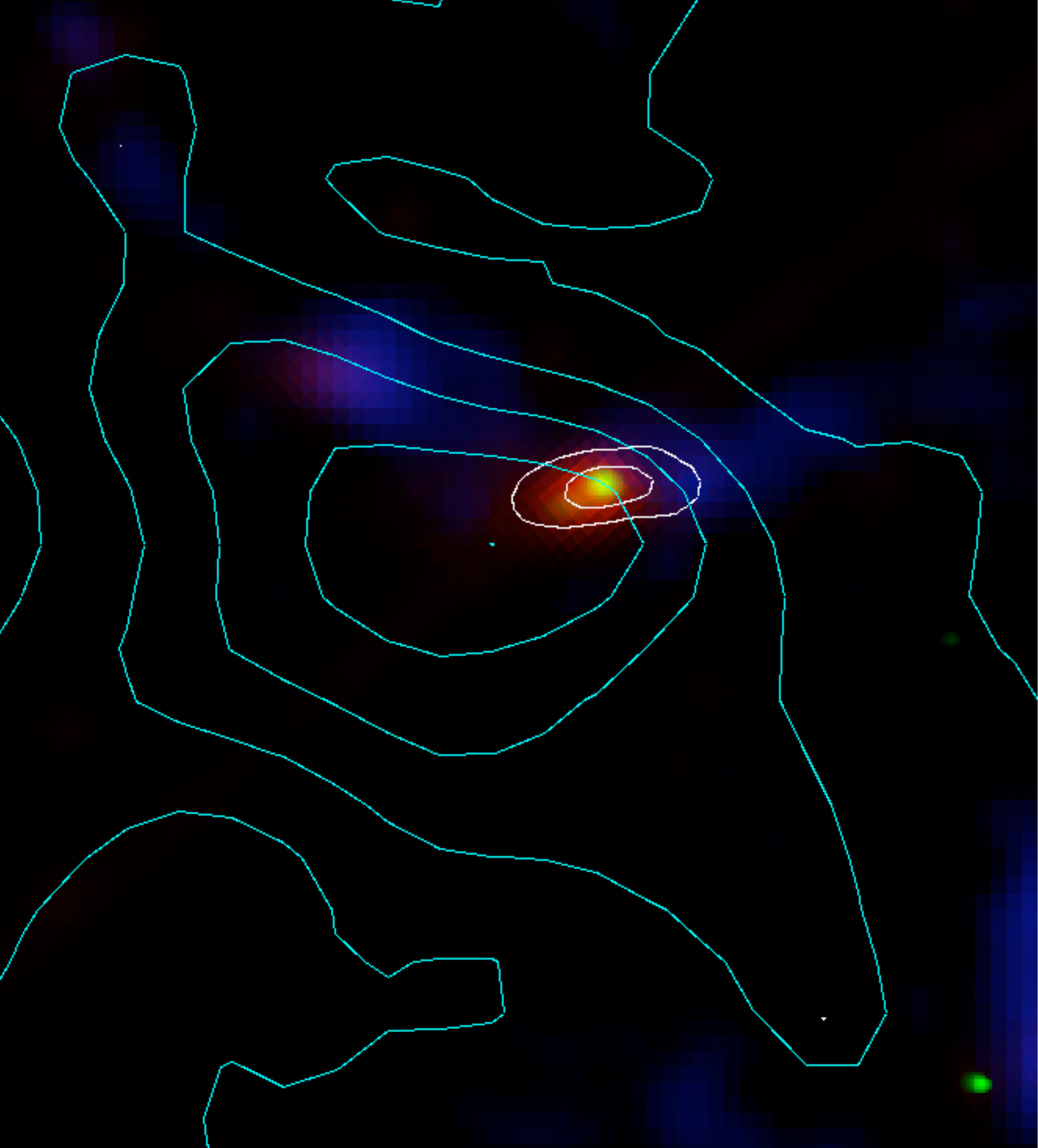}
\includegraphics[height=11cm]{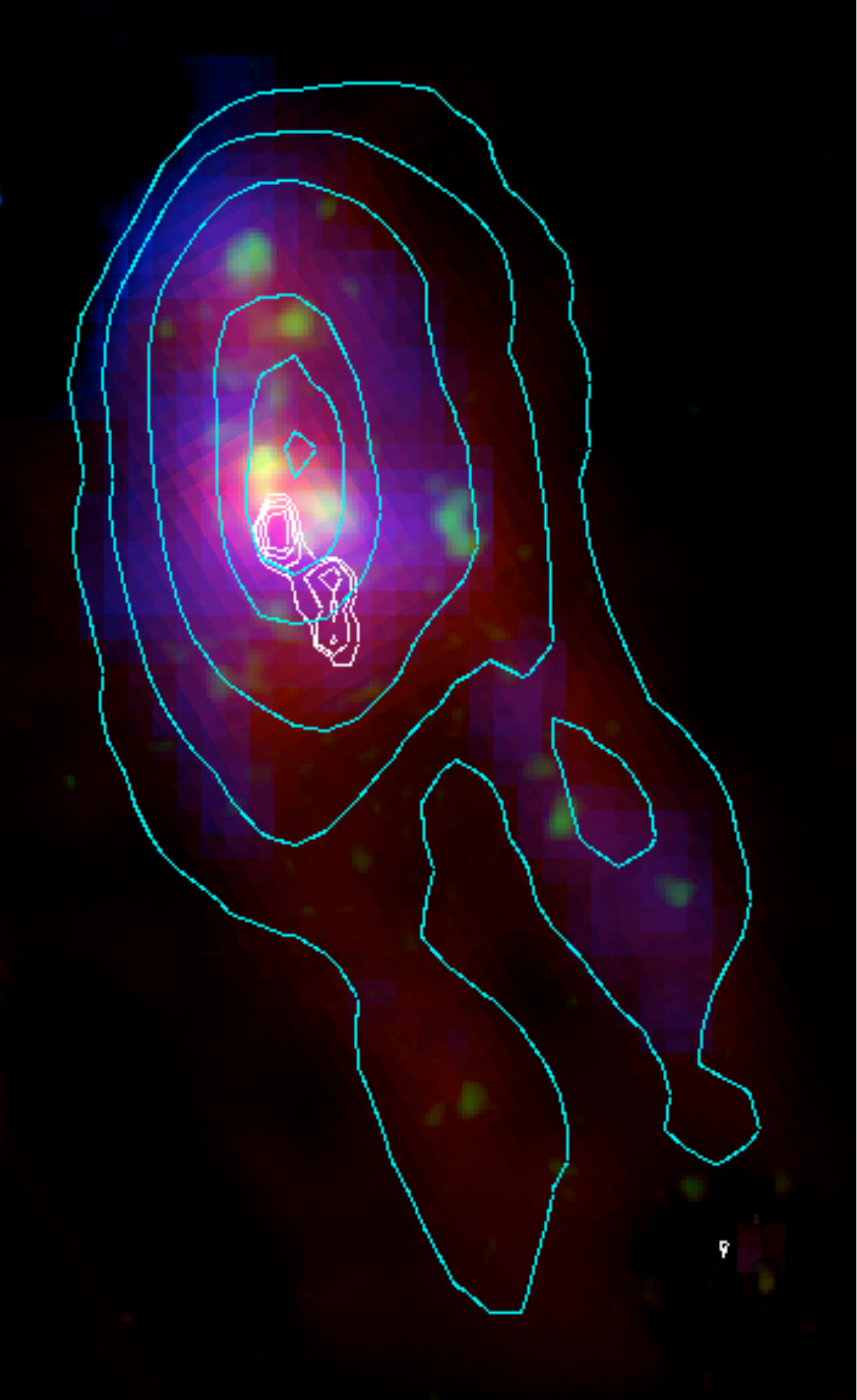}
\caption{Molecular gas in two TDG candidates: SQ-B, in the Stephan's Quintet (left) and NGC 2992 North (right). The CO(1-0) interferometric IRAM PdB and OVRO maps (white contours) are superimposed on a composite image showing various indicators of star-formation: far--UV emission from GALEX (blue), optical H$\alpha$ emission (green) and Spitzer/8 $\mu$m emission (red). The VLA/HI distribution is also shown with the  cyan contours. The image of SQ-B is adapted from Lisenfeld et al. (2004).} 
\label{fig1} 
\end{figure} 

Whether the lines from the carbon monoxide trace quite well, at least in metal-rich regions, the cool H$_{2}$, different tracers exist for the other phases of the molecular gas. In particular,  warm  H$_{2}$ may be directly detected in the mid--infrared. Higdon et al. (2006)  reported the detection of weak rotational  H$_{2}$ lines in the Spitzer/IRS spectrum of one of the SF regions in the collisional ring of NGC~5291.

\section{Probing cold molecular gas}

In addition to the regular, cool, H$_{2}$ as traced by CO, warm H$_{2}$ visible in the mid--infrared and shock heated H$_{2}$ (that emits  lines in the near-infrared, but has not yet been looked at in collisional debris), very cold H$_{2}$ has been speculated to exist (Pfenniger \& Combes 1994). It is not directly visible since it does not emit any specific lines, but being massive enough, its presence should affect the dynamics of galaxies. In any case, if originally present in the disk of galaxies, it should also end up in collisional debris. If the latter is gravitationally bound, one may determine its total, dynamical, mass, and comparing it with the luminous mass (consisting of HI, conventional H$_{2}$ and stars), infer a putative  missing mass presumably associated with this cold molecular gas. Note that this is the usual way the dark matter content of astrophysical objects is determined. In fact, collisional debris  doesn't accrete dark matter from cosmological {\em haloes}, which is  confirmed by numerical simulations (Barnes \& Hernquist 1992; Bournaud \& Duc 2006).
The method requires above all a very high precision in the determination of the various masses of the system, which so far could only be achieved in two somehow exceptional  systems presented here-below: 
NGC~5291 which contains prominent collisional debris, with numerous and active star-forming regions  (Bournaud et al. 2007) and the Virgo Tidal Dwarf candidate VCC~2062, which is close enough to allow a precise determination of its internal kinematics (Duc et al. 2007). Earlier investigations, in particular using H$\alpha$ and CO  data, had shown that Tidal Dwarf Galaxies had not dynamical to luminous mass ratios as high as ten like in normal dwarfs that are embedded in a dark matter halo  (Duc \& Mirabel,  1998; Braine et al. 2000). But these works lack spatial resolution  to more precisely compare the two masses.

\subsection{The giant collisional ring around NGC~5291}
NGC~5291 is an early--type galaxy located at the edge of a cluster of galaxies. It is surrounded by a prominent HI ring with a diameter exceeding 160 kpc and a mass as high as $5 \times 10^{10}~$M$_{\odot}$ (see Figure~2). Observations and the numerical model of  Bournaud et al. (2007)  favor a collisional origin, like for the famous Cartwheel galaxy: the ring was formed following the off-center  direct hit at a velocity of 1250 km~s$^{-1}$ of the gaseous disk of NGC~5291 by a massive intruder, 360 Myr ago. 
At several locations along the ring, the gas has collapsed and formed stars. About 30 of such star--forming regions have been identified thanks to their ultraviolet, H$\alpha$ and mid--infrared emission (Boquien et al. 2007). Molecular gas was detected in two of them (Braine et al. 2001). The internal kinematics of the three most prominent condensations could be studied  using Fabry-Perot H$\alpha$ and high-resolution VLA B--array HI datacubes. They exhibit flat rotation curves from which a dynamical mass has been derived. For the most massive object, it amounts to $30\pm8.6  \times 10^{8}~$M$_{\odot}$. The error of the method has been checked using numerical simulations. It turns out that all three objects have a total mass three times higher than the luminous one as inferred from 21~cm (HI), CO (tracing the H$_{2}$) assuming a galactic CO/H$_{2}$ conversion factor  and optical/near-infrared (tracing the stars) observations. The most natural candidate for the missing mass would be, in the CDM frame for non-baryonic dark matter,  a baryonic phase, likely under the form of cold molecular gas.

\begin{figure}[h]  
\centering 
\includegraphics[height=12cm]{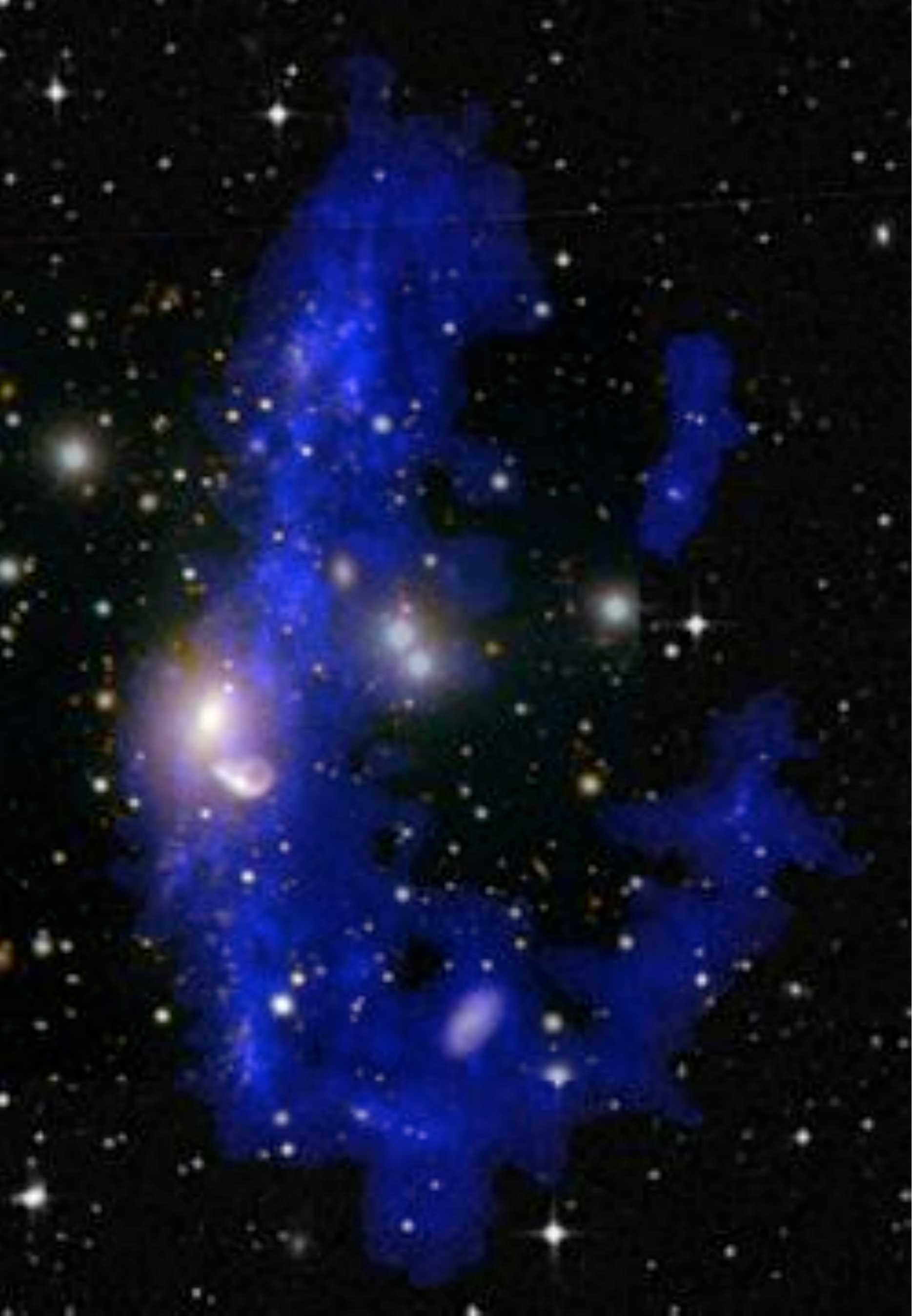}
\includegraphics[width=12cm,angle=90]{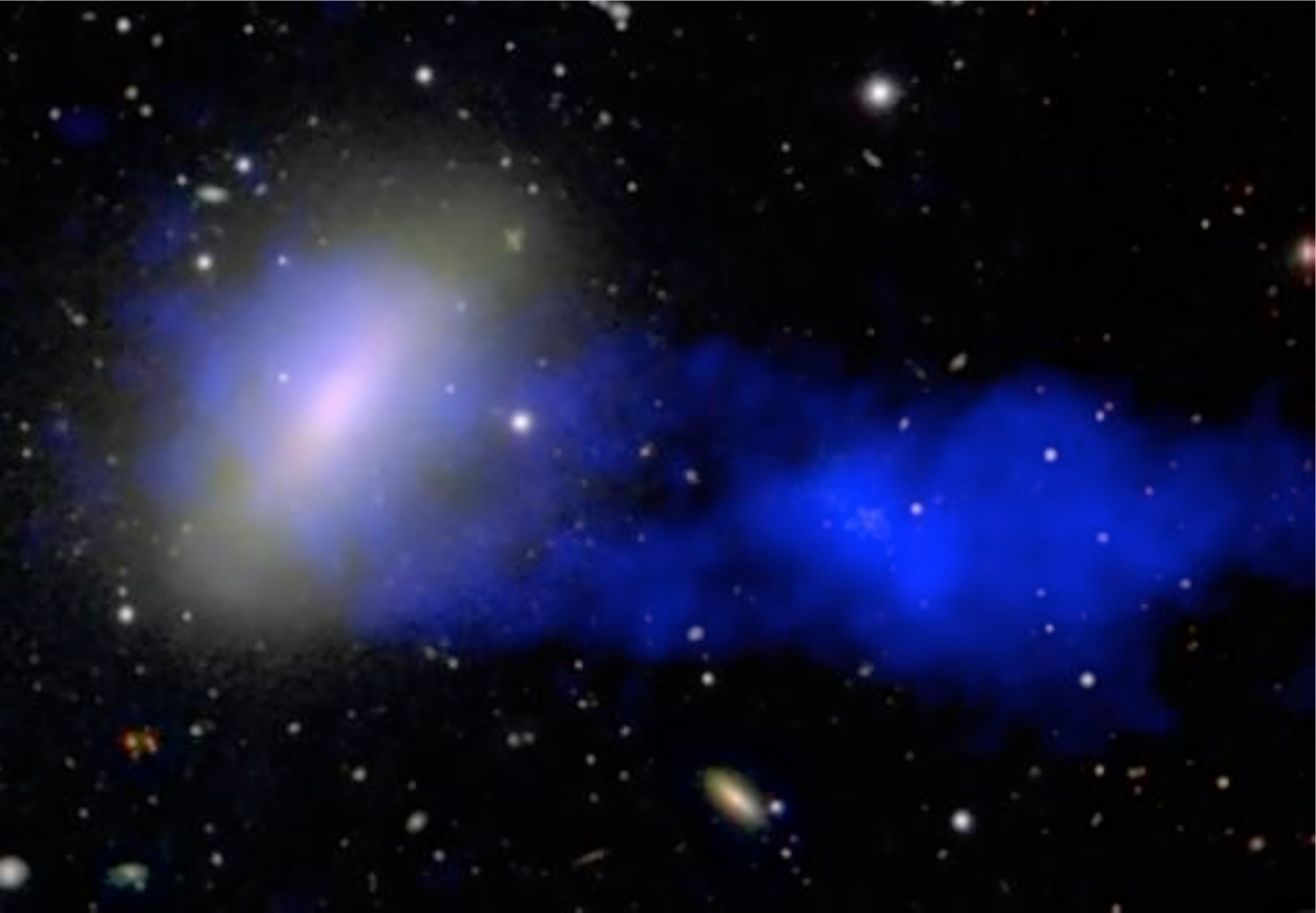}
\caption{The gaseous collisional debris around NGC 5291 (left) and NGC 4694/VCC 2062 (right). The VLA HI distribution is  superimposed in blue on true-color optical images. Adapted from Bournaud et al. (2007) and Duc et al. (2007).} 
\label{fig2} 
\end{figure}

\subsection{The nearby Tidal Dwarf Galaxy VCC~2062}

The Virgo Cluster hosts several cases of long HI filaments (Chung et al. 2007). One of the most massive ones is that emanating from the early--type galaxy NGC~4694, which we have studied in detail (Duc et al. 2007).  The recent VLA HI map from the VIVA project (Chung et al. 2005) is shown in Figure~2.

The tail, outside the optical body of  NGC~4694, contains about $10^9~$M$_{\odot}$ of HI. For most of its length, it has no optical counterpart. The exception is towards one of its most prominent condensations, where a low surface brightness stellar body is detected. It corresponds to the previously catalogued dwarf galaxy VCC~2062. There, the HI column density --- close to 10$^{21}$~cm$^{-2}$ --- is higher than the critical threshold for the on-set of star-formation. And, indeed, the dwarf exhibits star--forming regions as traced by H$\alpha$ and far--ultraviolet emission. Just to the South--West  of VCC~2062, another condensation has formed; its HI column density is even higher, but contrary to the previous object, it did not manage to  form stars and shows  at best an extremely faint optical counterpart. The CO(1-0) line was detected with the IRAM 30--m antenna towards both condensations; the flux is however higher towards the star--forming one. 
Why have these two nearby HI clouds such different properties? The answer probably comes from the kinematical analysis of the system. A Position--Velocity diagram based on the HI datacube reveals the presence of a well--defined velocity gradient (of about 45~km~s$^{-1}$ over 4~kpc) towards VCC~2062 while the SW condensation exhibits broad HI lines and no structured velocity field. In one case, the HI gas is kinematically decoupled from the rest of the tail, is gravitationally bound, rotating and forming stars like a normal galaxy, in the other it has not (yet) collapsed and has remained purely gaseous. 

Duc et al. (2007)  discuss various hypotheses for the origin of the HI--structure around VCC~2062. According to the most attractive one, the HI in the tail would come from a small HI--rich galaxy that has been accreted by NGC~4694 a few hundred Myr ago. In this scenario, the HI structure is a tidal tail and VCC~2062, which was formed in it, a Tidal Dwarf Galaxy. This is further suggested by its relatively high oxygen abundance of  about solar and the CO detection. 

The ratio between the dynamical mass of VCC~2062 --- 3-4$\times 10^{8}~$M$_{\odot}$,  as inferred from the velocity gradient of the HI --- and the luminous mass --- consisting of HI, H$_{2}$ traced by CO and stars --- estimated to be  only of order 2--3, is much lower than that found for field, dark--matter dominated  dwarf galaxies. This excludes the hypothesis that  VCC~2062 is the stellar remnant of a pre--existing  more massive galaxy which would have been tidally disrupted (but would not have yet been accreted by NGC~4694). On the other hand, the fact that the dynamical and luminous masses are not equal reveals in this object as well the presence of some missing matter and perhaps cold H$_{2}$.

\section{Summary and discussion}
Our detailed, multi-wavelength,  studies of collisional debris, especially its molecular gas content,  raise a number of questions.
\begin{itemize}

\item Where does the cool H$_{2}$ detected in quantities in tidal tails come from? Locally, out of tidally expelled, dense, HI clouds, as suggested by the kinematical data, and/or from the disk of the parent galaxy, as indicated by the detection of more diffuse CO emission outside the main centers of star-formation? The answer to this question will require to carry-out complete maps of the molecular gas along collisional debris. So far most of the CO detections result from pointed observations, biased towards regions of high HI column density.

\item How does the star-formation efficiency (the ratio of the star formation rate to the molecular mass) in collisional debris compare with that of individual star-forming regions of spiral disks and classical dwarf galaxies? A first rough comparison is presented in Braine et al. (2001) but should now be updated. The  SFE  should in particular be  determined in an homogeneous way, a difficult task for objects of different sizes and environments.  

\item Is really the missing mass found in collisional debris and hence in spiral disks very cold H$_{2}$? First of all, the  method used to determine the dynamical mass could be challenged. The objects formed in the debris are, spatially, barely resolved and  their inclination is not directly  determined;  however the uncertainties of the method have been well assessed by numerical simulations.
Besides, the conversion factor used to derive the mass of luminous  cool  H$_{2}$  from   the CO line intensity is not very constrained; in the relative high-Z environment of TDGs, it should however be not so different than the assumed galactic value.
 If real, the mismatch between the dynamical and the luminous mass, estimated to a factor of 2--3 in now 4 different objets,  can also be interpreted in a totally different manner. Cold streams of non-baryonic dark matter in disks is a theoretical possibility. 
  Gentile et al. (2007) and Milgrom (2007)  were able to reproduce the observed velocity curves of the recycled objects around NGC~5291 within the MOND framework.   On the other hand, new  simulations including conventional gravity, usual cosmological Dark-Matter halos and the putative cold component of H$_{2}$  in spiral disks (which was not implemented in the numerical modes of Bournaud et al.  2007) seem to also give a good match with the data (Revaz et al., in preparation).
 
\end{itemize}

 Although being just ``debris", tidal tails, rings and more in general all the material sent out of galaxies during their merging history, especially the molecular gas,  may tell about some fundamental aspects of astrophysics.

\begin{acknowledgements}
I wish to thank all my collaborators on this project, especially  Fr\'ed\'eric Bournaud, Jonathan Braine, Elias Brinks, Ute Lisenfeld and Fabian Walter  for their precious contribution on the detection of molecular gas in collisional debris. 
 
\end{acknowledgements} 







\end{document}